\documentclass[aps,prl,reprint,groupedaddress]{revtex4-2}

\usepackage{graphicx}
\usepackage{color}
\usepackage{caption}
\definecolor {myc} {rgb} {0,0,0}  %black, uncomment this line for regular copy
\usepackage{tikz}

\begin{document}

\newcommand{\vet}[1]{\mathbf{#1}}
\newcommand{\vect}[1]{\mathbf{\textit{#1}}}
\newcommand{\an}[1]{{\color{blue} \bf{#1}}}
\newcommand{\lc}[1]{{\color{red} \bf{XXX Luca's comment: #1}}}

%\preprint{}

\title{Role of normal stress in the creep dynamics and failure of a biopolymer gel}


\author{Angelo Pommella}
\affiliation{Laboratoire Charles Coulomb (L2C), Univ. Montpellier, CNRS, Montpellier, France}
\author{Luca Cipelletti}
\altaffiliation[Also at ]{Institut Universitaire de France (IUF)}
\affiliation{Laboratoire Charles Coulomb (L2C), Univ. Montpellier, CNRS, Montpellier, France}
\author{Laurence Ramos}
%\thanks{}
\affiliation{Laboratoire Charles Coulomb (L2C), Univ. Montpellier, CNRS, Montpellier, France}


\date{\today}

\begin{abstract}
We investigate the delayed rupture of biopolymer gels under a constant shear load by simultaneous dynamic light scattering and rheology measurements. We unveil the crucial role of normal stresses built up during gelation: all samples that eventually fracture self-weaken during the gelation process, as revealed by a partial relaxation of the normal stress concomitant to a burst of microscopic plastic rearrangements. Upon applying a shear stress, weakened gels exhibit in the creep regime distinctive signatures in their microscopic dynamics, which anticipate macroscopic fracture by up to thousands of seconds. The dynamics in fracturing gels are faster than those of non-fracturing gels and exhibit large spatio-temporal fluctuations. A spatially localized region with significant plasticity eventually nucleates, expands progressively, and finally invades the whole sample triggering macroscopic failure.
\end{abstract}

\pacs{}
%\keywords{}

\maketitle

Failure in soft materials is important from scientific and practical perspectives~\cite{Creton16}. Understanding how and when failure may occur is crucial to design materials that are able to prevent or, conversely, to facilitate rupture, depending on applications. Furthermore, why failure may occur after a long induction time is a fascinating question, crucial in fields as geology, economics, and medicine~\cite{cipelletti_microscopic_2019}. In soft matter, many key questions remain unanswered in spite of recent progress in understanding failure phenomena in polymers~\cite{Kinloch13}, granular materials~\cite{Tapia16,Schmeink17} and networks formed by colloids~\cite{Smith10,Aime18}, and biopolymers~\cite{Bonn98,Baumberger06,Baumberger2006,barrangou_textural_2006,Buehler07,daniels_instabilities_2007,spandagos_surface_2012,Leocmach14,mao_heat-induced_2017}.

Biopolymer gels are very relevant and largely exploited in drug delivery, personal care, and tissue engineering, thanks to remarkable properties such as high water content and softness. Another distinctive feature of biopolymers {\color {myc} forming fibrils} is their ability to develop a negative normal stress under shear, due to the stiffness of the {\color {myc}fibrils}~\cite{Janmey06,Licup15,deCagny16,Yamamoto17,Baumgarten18}. This property is responsible for strain hardening, which allows the gels to stiffen under a large load to prevent rupture, while preserving flexibility at smaller loads. Despite the relevance of these materials, little is known on the microscopic mechanisms leading to failure and their interplay with the distinctive mechanical features of biopolymer gels. Indeed, most studies have focused on crack nucleation and propagation at the macroscopic level~\cite{Bonn98,Kong03,Baumberger2006}, rather than on the evolution of the microscopic structure and dynamics before failure.

We address these questions by coupling dynamic light scattering to rheology to rationalize the failure process of biopolymer gels. We demonstrate the crucial role of normal stresses developed during gelation for the ultimate behavior of the gel under a constant shear load, showing that gels that eventually fail have suffered irreversible damages due to tensile stress self-generated during gelation. We propose that these irreversible rearrangements modify the bond distribution in the sample, ultimately diminishing its resistance to shear stress. Finally, we unveil several kinds of dynamic precursors of failure, highlighting the crucial role of microscopic dynamics for understanding failure in soft materials.

\begin{figure}
\includegraphics[width=8.5 cm]{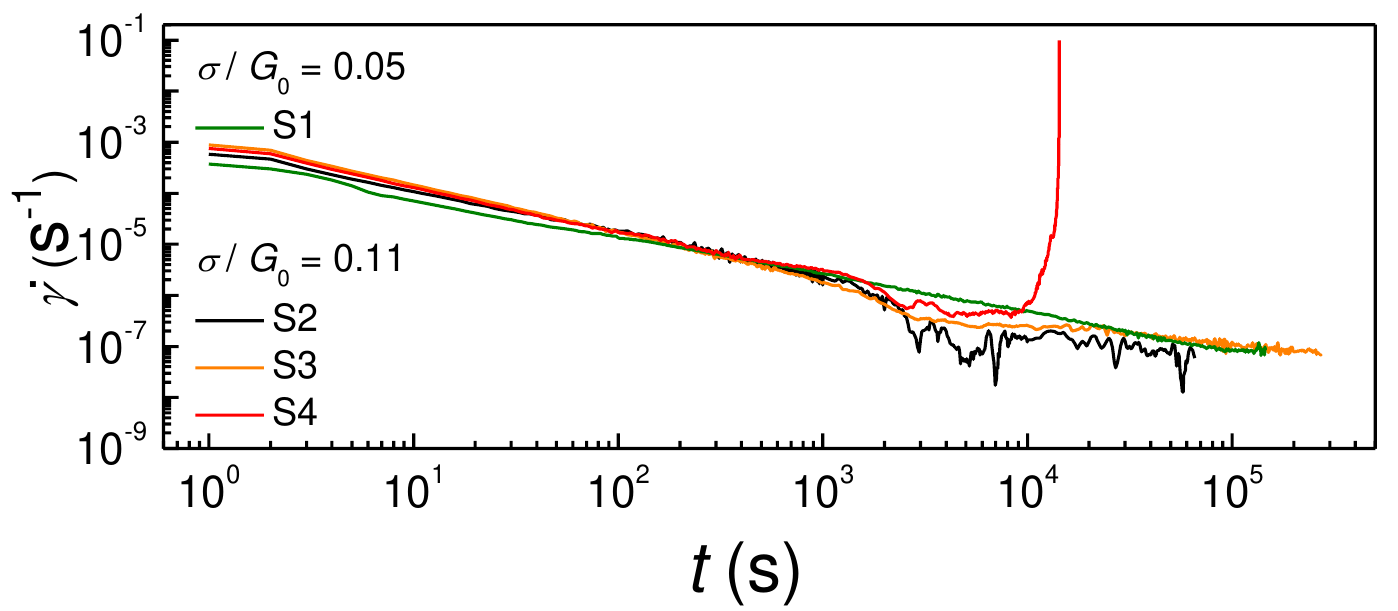}
\caption{\label{Fig1} Shear rate \textit{vs} time during creep tests, for samples S1-S4. $\sigma$ is the amplitude of the applied stress, $G_0$ the gel elastic plateau modulus.}
\end{figure}

We perform creep tests on agarose gels~\cite{Heymann35, payens_transition_1972, Clark1987}, using a rheometer coupled to a custom-made light scattering apparatus~\cite{pommella_coupling_2019}. Gelation is induced in-situ, by cooling hot agarose solutions to $23^{\circ}$C {\color {myc} yielding a network made of aggregated fibrils of double helices, with a typical mesh size of $100$ nm~\cite{Djabourov}} (details in~\cite{SM1}). Creep experiments start $24$ hours after gelation to allow for equilibration of the gel; during creep, the time evolution of the strain $\gamma$ is measured for up to $24$ hours. Figure~\ref{Fig1} shows the strain rate $\dot{\gamma}$ \textit{vs} $t$, the time elapsed since the application of a shear stress step $\sigma$, for four samples with the same composition and comparable viscoelasticity~\footnote{The elastic modulus $G_0$ of the samples is shown in Fig. 5b and will be discussed later.}. In all cases, after an initial elastic jump at $t=0$, $\dot{\gamma}$ decreases as $\dot{\gamma} \propto t^a$, $a = -0.83 \pm 0.10$. For sample S4, an upturn of $\dot{\gamma}(t)$ occurs at $t \approx 10^4$ s, followed by a fast increase of $\dot{\gamma}$, along with the full relaxation of the normal stress (Fig.~SM3~\cite{SM1}), signing the gel macroscopic failure. Remarkably, failure only occurs for S4, although all samples have comparable moduli: for S1-S3, $\dot{\gamma}$ keeps decreasing with the same power law, up to waiting times $10$ times longer than the fail time of S4. As expected intuitively $\sigma$ must be sufficiently large for macroscopic failure to occur within the duration of the creep measurement ($\geq 24$ h). Intriguingly however, samples can ultimately display drastically different behaviors under a similar load, despite exhibiting the same creep response over several hours.

To rationalize the findings of Fig.~\ref{Fig1}, we couple light scattering to rheometry. The microscopic dynamics are probed with space and time resolution by acquiring a time series of speckle patterns backscattered by the sample. The images are processed to calculate $c_I (t, \tau, \vet{r})$, a local, two-time degree of correlation~\cite{duri_resolving_2009,cipelletti_simultaneous_2013} with $t$ the time, $\tau$ a time delay, and $\vet{r}$ the position in the sample, $\vet{r}=0$ being the coordinate of the rheometer axis of rotation~\cite{SM1}. $c_I$ quantifies the amount of microscopic motion over the lag $\tau$, averaged over the sample thickness: $c_I \rightarrow 0$ for motion over distances $1/q \gtrsim 30$ nm, with $q=33.2~\mu \mathrm{m}^{-1}$ the magnitude of the scattering vector~\cite{pommella_coupling_2019}.

\begin{figure}
\includegraphics[width=8 cm]{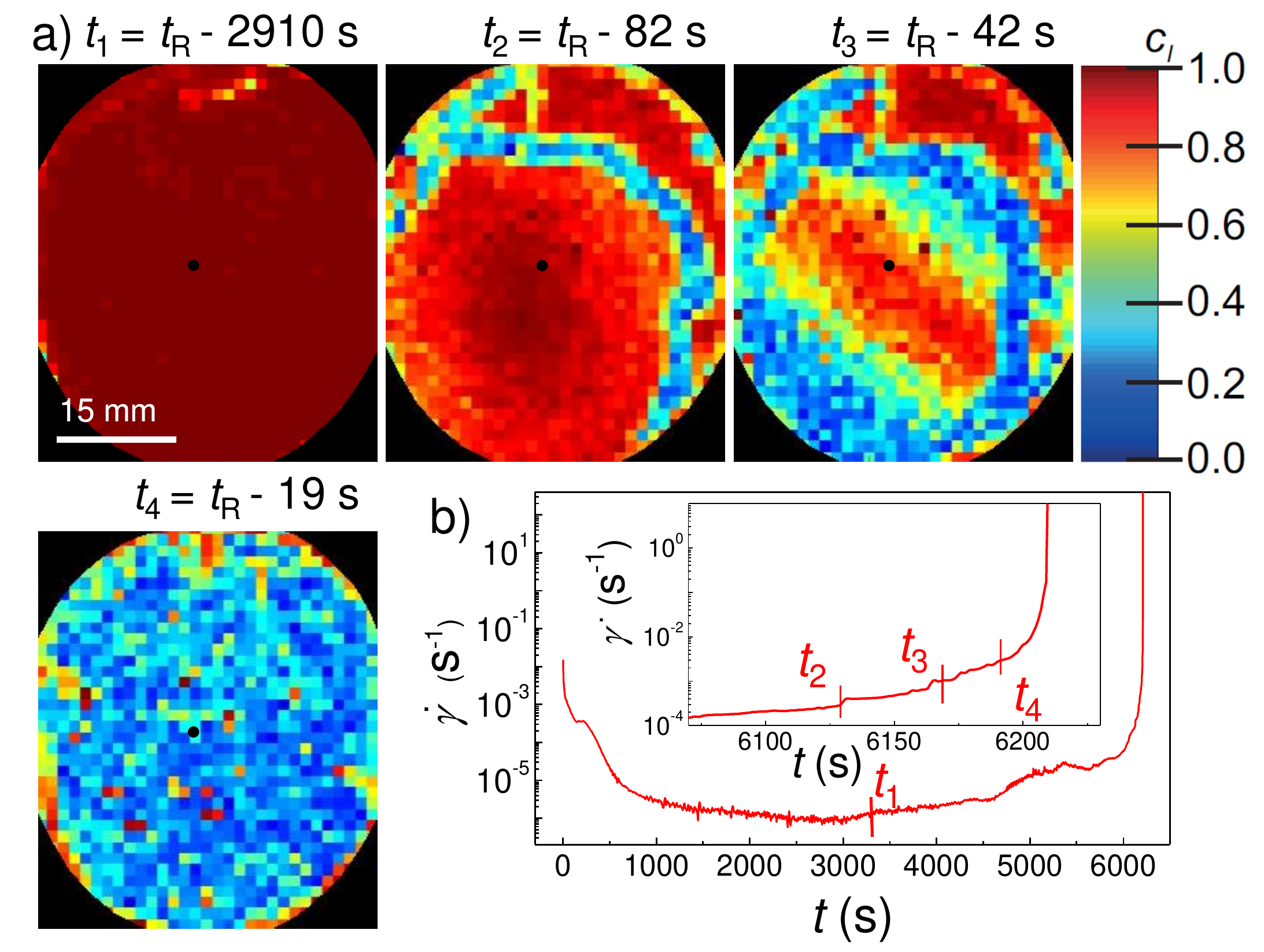}
\caption{\label{Fig2} (a) Dynamic activity maps quantifying the dynamics over a time delay $\tau=1$ s, measured at four times before rupture, at $t_R = 6210~\mathrm{s}$. The black points indicate the plate rotation axis. (b) Creep curve showing $t_1$ (main plot) and $t_2-t_4$ (inset).}
\end{figure}

We show in Fig.~\ref{Fig2}a, for a sample of comparable viscoelasticity as those of Fig.~\ref{Fig1}, a dynamic activity map (DAM)~\cite{duri_resolving_2009} obtained by dividing the imaged area (here almost the whole sample, see Fig. SM2b in~\cite{SM1}) into square ROIs of side $1.2~\rm{mm}$, and plotting for each ROI $c_I$ at a fixed time lag $\tau =1$ s, with colors ranging from dark red (frozen dynamics, $c_I=1$) to blue (intense dynamic activity, $c_I < 0.01$, corresponding to a mean square displacement of the scatterers $\Delta r^2(\tau=1 s) > 1.4 \times 10^{-2}~\mu\mathrm{m}^2$). The DAMs in Fig.~\ref{Fig2}a illustrate the gel dynamics at different times before rupture at $t = t_R=6210~\mathrm{s}$ (see also the movie in~\cite{SM1}). Up to $t_1 = t_R-2910$ s, approximately at the minimum of the macroscopic shear rate, the dynamics are almost frozen, except for a small region (top of the DAM). Upon approaching failure, this region first expands along the sample rim, then propagates inward (see DAMs at $t_2 = t_R-82~\mathrm{s}$ and $t_3 = t_R-42~\mathrm{s}$), until the entire gel exhibits intense dynamics (at $t_4 = t_R-19$ s). At $t_4$, the speckle pattern
 shows radially propagating cracks in the bulk of the gel (see Fig. SM6 in~\cite{SM1}). The inward propagation of intense dynamics is likely due to the heterogeneous strain profile in the plate-plate geometry, for which the local strain is proportional to $r$. Note however that the DAM at the onset of plasticity does not exhibit the same circular symmetry as the strain field.  Fast dynamics nucleate in a specific region near the rim, presumably where the gel is weaker, due to heterogeneities in its structure. The nucleation of this dynamically active region about $3000$ s before rupture constitutes a remarkable dynamic precursor of failure {\color {myc} (see movies and Fig.~SM7 in~\cite{SM1} for DAMs from other experiments)}.

The DAMs of Fig.~\ref{Fig2} highlight the rearrangement dynamics on short timescales, $\tau = 1~\mathrm{s}$. To explore the timescale dependence of the dynamics, we compute $c_I$ at various $\tau$ during the creep tests of samples S1-S4 shown in Fig.~\ref{Fig1}, zooming into a small area of size $2\times1 \rm{mm}^2$ (see Fig. SM2c in~\cite{SM1}). The region is chosen so as to minimize the trivial contribution to $c_I$ arising from the affine deformation of the sample during creep~\cite{SM1}. Data for the four samples are show in Fig. SM5 in~\cite{SM1}. Figure~\ref{Fig3}a focuses on sample S4, during the last $900$ s before failure, at $t_R=11825~\mathrm{s}$. The drops of $c_I$ reflect plastic rearrangements leading to enhanced microscopic dynamics. For $\tau = 2~\mathrm{s}$,  $c_I(t,\tau)$ exhibits an abrupt drop at $t \approx 11500$ s, a few hundreds of seconds before the gel rupture. This sudden drop, followed by a slow recovery, is consistent with a wave of plastic activity, highly localized in time, sweeping the sample, as in Fig.~\ref{Fig2}a. $c_I$ traces for longer delays, $\tau = 10$ and $40~\mathrm{s}$, reveal an additional burst of plasticity around $t = 11100~\mathrm{s}$, {followed by a progressive acceleration of the dynamics, undetectable when probing the dynamics for $\tau = 2~\mathrm{s}$. This constitutes a second dynamic precursor of failure, independent of the first one, based on the evolution of the dynamics on longer timescales.

A third dynamic precursor of failure clearly emerges when plotting $g_2(\tau)-1$, the $t$-averaged degree of correlation shown in Fig.~\ref{Fig3}b~\footnote{The time average is performed over the duration of the creep experiment, excluding the first 1000~s, when the dynamics are faster because of the stress step imposed at $t=0$. For S4, we also exclude the last 1000 s before failure, when the dynamics accelerate as discussed above.}. Correlation functions for all samples can be fitted with a generalized exponential decay, $g_2 (\tau) - 1 = \exp [-(\tau/\tau_c)^p]$. For non-rupturing samples (all curves in Fig.~\ref{Fig3}b, except S4), the relaxation time $\tau_c$ lays in the range $(3000-8000)$ s~\cite{SM1}, comparable to that of a gel at rest (blue squares), suggesting that the microscopic dynamics are not significantly perturbed by the imposed stress. By contrast, for S4, the gel that eventually fails, $\tau_c = 140$ s, more than $20$ times shorter than for non-rupturing gels.

\begin{figure}
\includegraphics[width=8 cm]{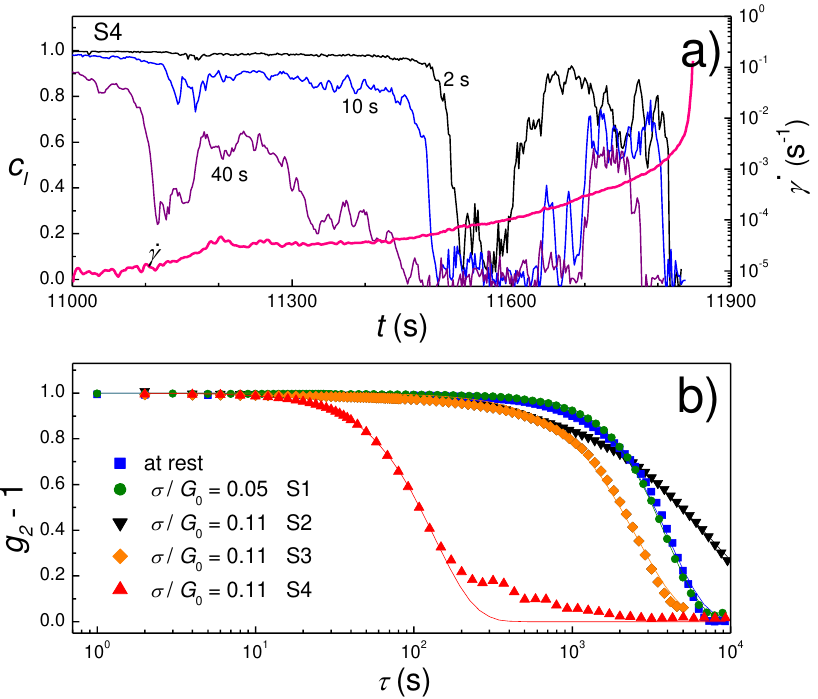}
\caption{\label{Fig3} (a) Time evolution of the degree of correlation at different delays $\tau$ for S4, the sample that will fail. Only the last $900$ s before rupture are shown. Thick pink line: shear rate $\dot{\gamma}$. (b) Symbols: intensity correlation functions for the same samples as in Fig.~\ref{Fig1}. Lines: generalized exponential fits (see text).}
\end{figure}

Figure~\ref{Fig3}b demonstrates that the ultimate fate of a gel is encoded in its average dynamics during the whole creep experiment, not just on approaching failure. Thus, failing gels must differ from non-rupturing ones since their very formation, {\color {myc} as inferred from experiments on colloidal gels although only with macroscopic quantities~\cite{sprakel_stress_2011,helal_simultaneous_2016}}. However, any differences must here be subtle: conventional rheological quantities such as the storage, $G'$, and loss, $G"$, viscoelastic moduli do not allow one to identify beforehand those samples that will fail. Indeed, gels with the same composition consistently have a very similar dependence of linear viscoelasticity on time $t_q$ since starting the cooling ramp to induce gelation. This is shown by the symbols of Fig.~\ref{Fig4}a, which display the time evolution of $G'$ and $G''$ for three distinct gels with the same composition as those of Figs.~\ref{Fig1} and~\ref{Fig3}. By contrast, the time dependence of the normal stress $\sigma_N$ (lines) exhibits significant sample-to-sample differences. The three lines in Fig.~\ref{Fig4}a are representative of three distinct classes of behavior, based on the evolution of $\sigma_N$ during gelation and on how the gap $H$ is controlled~\cite{mao_normal_2016}. When $H$ is kept constant (solid and dotted lines in Fig.~\ref{Fig4}a), a negative $\sigma_N$ signals the emergence of normal forces that pull the plates together. These forces may relax during gelation, leading to gels with a moderate normal stress plateau $\sigma_N^0$ at the end of gelation ($\sigma_N^0\sim -0.8$ to $-5~\mathrm{kPa}$, continuous line in Fig.~\ref{Fig4}a). If no relaxation occurs during gelation, $\sigma_N^0 < -5~\mathrm{kPa}$, (dotted line). We also prepare stress-free samples, obtained by letting the rheometer continuously adjust $H$ to keep $\sigma_N\approx 0$ (dashed line in Fig.~\ref{Fig4}a).

To understand the microscopic mechanisms responsible for the distinct evolutions of $\sigma_N$, we plot in Figs.~\ref{Fig4}b,c the relevant macroscopic quantities together with the evolution of $c_I$ at selected delays $\tau$. Under $\sigma_N \approx 0 $ conditions (Fig.~\ref{Fig4}b), $H$ decreases rapidly at the onset of gelation, due to the emergence of internal tensile forces. This macroscopic reorganization is mirrored by a transient acceleration of the microscopic gel dynamics, signalled by a drop of $c_I$ that lasts until $t_q \sim 900-1500~\mathrm{s}$, depending on $\tau$. As gelation proceeds, $H$ tends to stabilize, suggesting less active rearrangements, as confirmed by the $c_I$ traces, which for $t_q \gtrsim 2500~\mathrm{s}$ reach values close to one, indicative of frozen dynamics. Only sporadic, limited losses of correlation are seen, similar to the `quakes' reported in a variety of gel systems and ascribed to the relaxation of residual internal stresses built up during gelation~\cite{Cipelletti00,filiberti_multiscale_2019}.

The scenario is very different when gelation is performed at fixed $H$ (Fig.~\ref{Fig4}c): a negative $\sigma_N$ develops as gelation starts; after reaching a minimum (here, $\sim -6$ kPa), $\sigma_N$ relaxes approaching a plateau $\sigma_N^0$ (here, $\sim-2.5$ kPa). The upturn of $\sigma_N$ is concomitant with a strong, transient acceleration of the microscopic dynamics (non-monotonic evolution of $c_I$ for $1600~\mathrm{s} \lesssim t_q \lesssim 4000~\mathrm{s}$). Note that the acceleration of the dynamics is much more pronounced here than for the sample with variable $H$ and $\sigma_N \approx 0$, since comparable or even greater drops of $c_I$ are seen over time lags $\tau$ significantly smaller than in Fig.~\ref{Fig4}b.
We conclude that rearrangements during the late stages of gel formation, triggered by the development of strong normal forces, are responsible for the remodeling of the gel and the partial relaxation of $\sigma_N$. These rearrangements do not systematically occur in all experiments at fixed $H$, suggesting random sample-to-sample variations of the gel ability to sustain large normal stresses. Accordingly, for some samples $\sigma_N$ decreases monotonically before reaching a plateau (dotted line in Fig.~\ref{Fig4}a).

\begin{figure}
\includegraphics[width=7.5 cm]{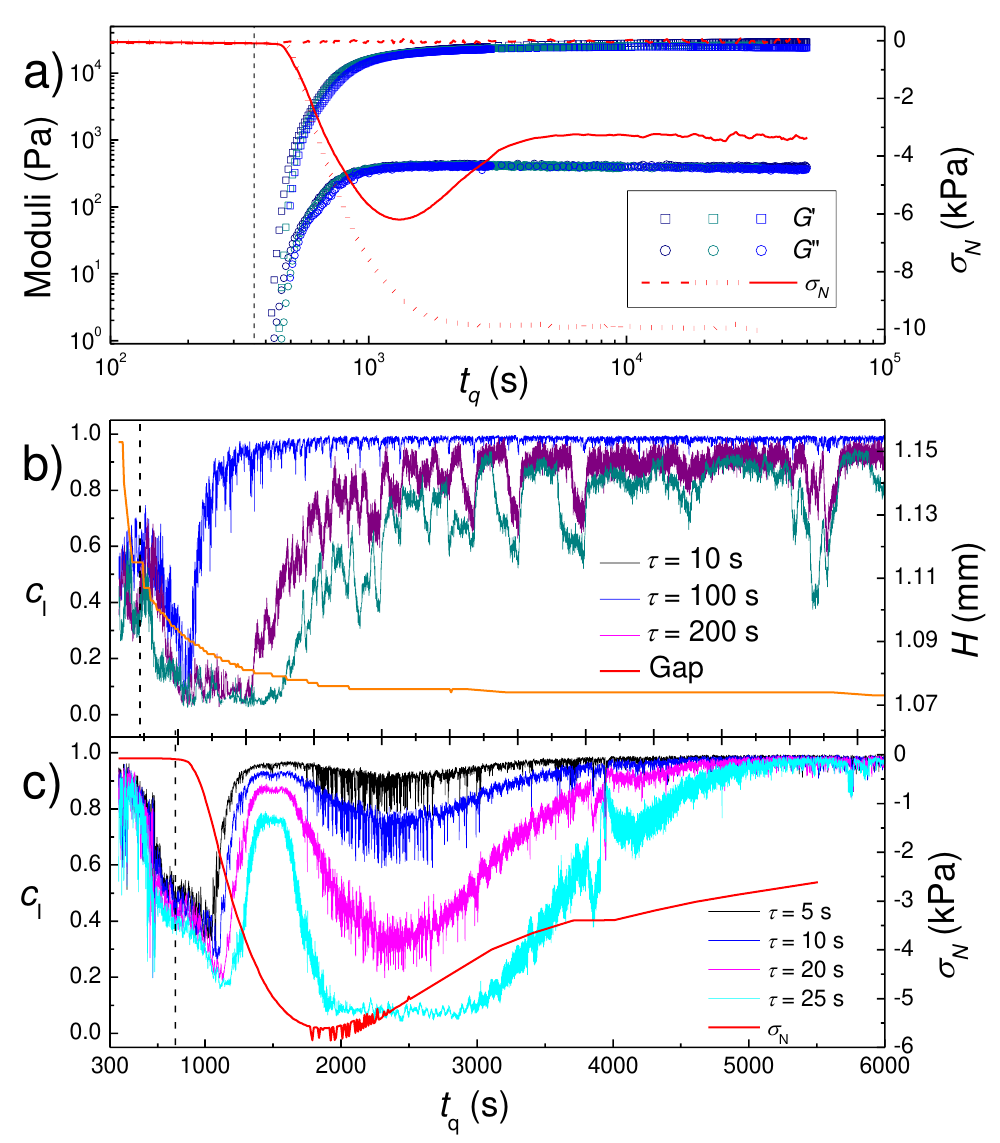}
\caption{\label{Fig4} Evolution with time $t_q$ since quenching the sample at $T=23^{\circ}$C, of various quantities. (a) Viscoelastic moduli $G'$ and $G''$ (symbols) and normal stress (lines), for three samples during gelation, at fixed gap $H$ (solid and dotted lines) or adjustable $H$ and $\sigma_N \approx 0$ (dashed line). Vertical line: time at which $T$ reaches $23^{\circ}$C. (b), (c): Degree of correlation during gelation at variable (b) and fixed (c) gap. In (b), the orange line is the gap evolution. In (c), the red line is $\sigma_N(t)$. The vertical lines in b), c) indicate $G'=G"$.}
\end{figure}

To rationalize the role of the normal stress on the fate of sheared gels, we design a protocol that reproduces the non-monotonic behavior of $\sigma_N$ in a controlled fashion. We induce a change of $\sigma_N$ through incremental variations of the gap, starting from a relaxed gel ($\sigma_N^0 \approx 0$), and recording at each step both $\sigma_N$ and the elastic modulus $G_0$, measured by imposing an oscillatory shear of small amplitude. Figure~\ref{Fig5}a shows the sample trajectory in the parameter space $(H,|\sigma_N|, G_0)$; relevant points along the trajectory are labeled \textcircled{1}---\textcircled{4}. Starting from \textcircled{1} and increasing $H$, $|\sigma_N|$ increases progressively, inducing a smooth growth of $G_0$, as previously observed for biopolymers~\cite{Licup15,mao_normal_2016}. As seen by projecting the trajectory on the $(H,|\sigma_N|)$ plane (red line and symbols), beyond point \textcircled{2} $\sigma_N$ stops growing with $H$ and reaches a plateau, up to \textcircled{3}. This is the distinctive signature of plastic damage due to the tensile strain imposed to the gel, as confirmed by the observation of hysteresis: upon reducing $H$, from \textcircled{3} to \textcircled{4}, the $\sigma_N = 0$ condition is recovered for a gap $H$ larger than the initial one. The evolution of $G_0$ is strikingly different from that of $\sigma_N$ (blue lines and symbols, Fig.~\ref{Fig5}a). Between \textcircled{2} and \textcircled{3}, the shear modulus \textit{steeply increases}, rather than flattening out. This strain hardening is irreversible: when $H$ is reduced to recover $\sigma_N = 0$ in \textcircled{4}, $G_0$ remains significantly larger than in \textcircled{1}.

Figure~\ref{Fig5}a suggests that as $\sigma_N$ becomes too large, bonds sustaining forces along the normal direction are broken, while new bonds are formed in other directions. This anisotropy leads to a gel with a shear modulus significantly higher than in pristine gels, but weaker under a tensile stress. We propose that this mechanism is responsible for the failure of our gels under creep. Both the non-monotonic $\sigma_N(t)$ and the microscopic dynamics shown in Fig.~\ref{Fig4} support the occurrence of these plastic rearrangements. As a result, some gels are significantly weakened in their ability to resist normal stresses. Although these gels would have larger $G_0$, they will paradoxically fail under shear, because biopolymer gels under shear quite generally develop strong tensile stresses~\cite{Janmey06,Licup15,deCagny16,Yamamoto17,Baumgarten18}.

We test this hypothesis by analyzing the behavior of a large number of gels with identical composition in creep tests lasting up to one day, focusing on the value of the normal stress plateau $\sigma_N^0$ and the shear modulus $G_0$ at the end of gelation. As shown by the black squares in Fig.~\ref{Fig5}b, gels able to sustain large normal stresses ($\sigma_N^0/\sigma < -3$), or which relaxed during gelation ending in a nearly stress-free configuration ($\sigma_N^0/\sigma \gtrsim -1.5$) are able to sustain the extra normal stress developed during creep. Conversely, gels with intermediate $\sigma_N^0$, comparable to $\sigma$, are susceptible to fail. Among them, all gels that actually failed (red circles) had a non-monotonic evolution of $\sigma_N$ and $G_0$ higher than the typical value of the other gels {\color {myc} ($G_0$ is about $50$ \% higher than for the reference gel, with $\sigma_N^0=0$, which does not fail)}~\footnote{{\color {myc} The loss moduli are also higher for the gels that fail than for the other gels, see Fig. SM8 in [28]}}. This finding demonstrate that the gels that fail underwent rearrangement processes breaking the isotropic distribution of the agarose chains in the gel network. These rearrangements do not significantly change the average structure of the gel, since the average static scattered intensity remains constant. However, they are unambiguously detected from changes in the gel dynamics.

\begin{figure}
\includegraphics[width=8 cm]{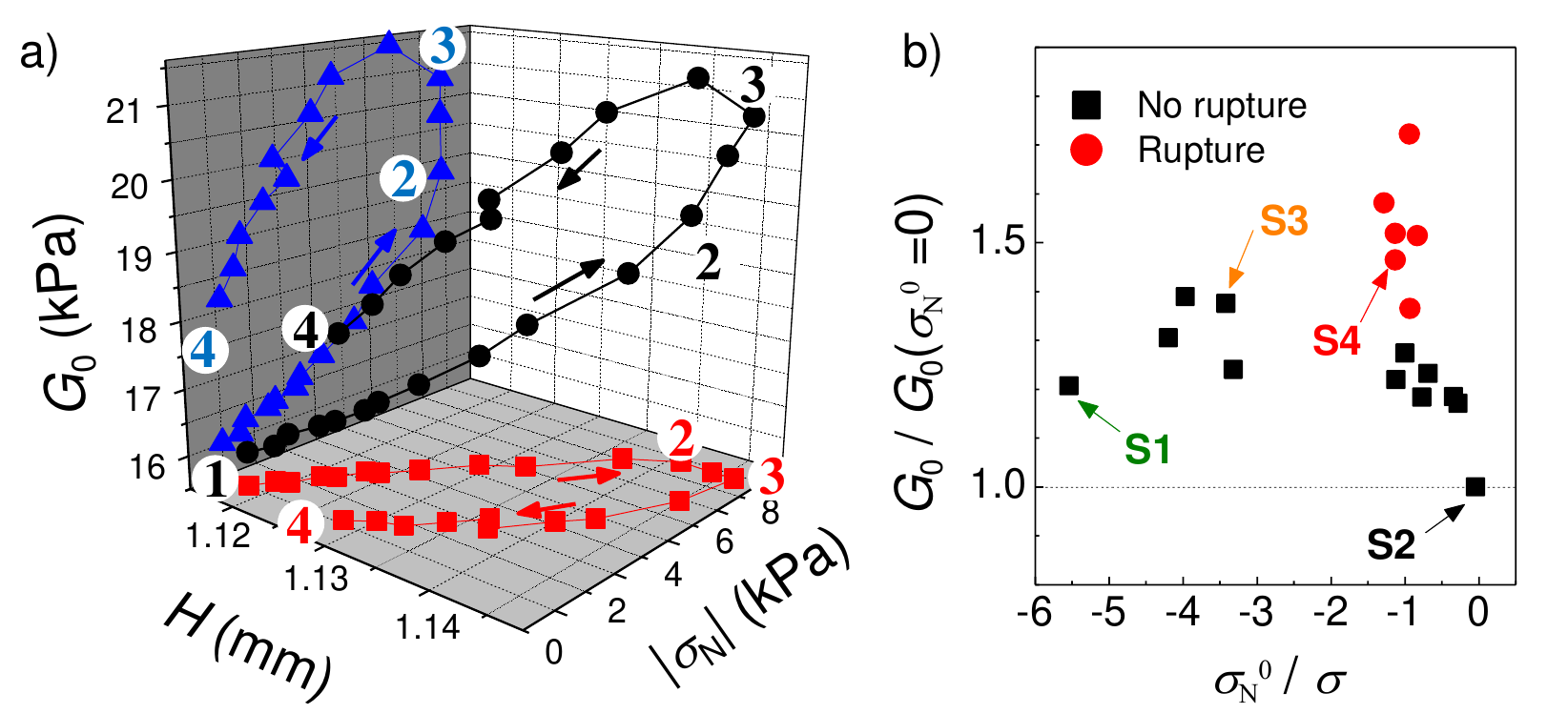}
\caption{\label{Fig5}(a) Black line and symbols: sample trajectory in the $(H,|\sigma_N|, G_0)$ parameter space for an experiment where the gap $H$ is varied. The arrows and numbers indicate the progression along the trajectory. Red, resp. blue, lines and symbols are the projection of the trajectory on the ($H$, $\sigma_N$), resp. ($G_0$, $\sigma_N$), plane. (b) $G_0$ {\color {myc}normalized by $17.8$ kPa, $G_0$ for sample (S2) with $\sigma_N^0=0$}, as a function of $\sigma_N^0/\sigma$ for gels that eventually did (red circles) or did not (black squares) fail in creep tests. The arrows indicate samples S1-S4 investigated in Figs.~\ref{Fig1} and~\ref{Fig3}.}
\end{figure}

In conclusion, we have unveiled the crucial role of normal stress for the ultimate fate of agarose biopolymer gels under a constant shear load, showing that failure is not related to lack of resistance in the shear direction, but rather to weakness under the tensile stress that accompanies shear in these materials. As recently observed for other soft solids~\cite{cipelletti_microscopic_2019}, macroscopic failure is preceded by several precursors in the microscopic dynamics. These findings pave the way for both anticipating and controlling the non-linear behavior of biopolymer gels, an important class of soft networks.

\begin{acknowledgments}
We acknowledge financial support from the French ANR (grant No. ANR-14-CE32-0005, FAPRES) and CNES. We thank E. Del Gado for fruitful discussions. LC acknowledges support from the Institut Universitaire de France.

L.C. and L.R. contributed equally to this work.
\end{acknowledgments}

\newpage
\begin{widetext}

\appendix
\textbf{Supplemental material to `Role of normal stress in the creep dynamics and failure of a biopolymer gel'}

We provide here (1) details on the sample preparation and the experimental set-up, (2) details on the computation of the two-time degree of correlation and of the affine contribution to the degree of correlation during creep experiments, and (3) additional mechanical and light scattering data from experiments on the same samples as in the main manuscript.

\section{Sample preparation and experimental set-up}

\textit{Samples.} The gels are prepared by mixing agarose powder (Sigma Aldrich A9539-10G, 1\% by weight) with MilliQ water at $95^{\circ}$C for $15$ minutes. The hot solution is then poured between the glass plates (diameter $50$ mm) of the rheometer that are pre-heated at $80^{\circ}$C.  After setting the gap $H$ between the plates to $H \approx 1$ mm and painting the sample rim with silicon oil to prevent water evaporation, the temperature is cooled down to $23^{\circ}$C to form the gel, at an imposed rate of $- 10^{\circ}$C/min. Gelation is monitored by applying an oscillating shear strain at frequency $10~\mathrm{rad~s}^{-1}$ and very low strain amplitude, $\gamma = 0.03\%$, and by recording the time evolution of the elastic, $G'$, and loss, $G''$, moduli.  The sample is a network made of aggregated fibrils of double helices held by H-bonds, with fibrils thickness of $3$ and $9$ nm. For a gel with $1$ \% agarose, the typical pore size is of the order of $100$ nm~\cite{Djabourov}.

To prevent sample debonding and slippage, the plates are treated with polyethyleneimine, a cationic polymer that promotes the anchoring of the anionic agarose chains to the negatively-charged glass plates~\cite{Christel12, Poptoshev02}. Figure SM1 shows typical pictures of a broken gel, taken at the end of a creep test, when lifting the rheometer upper plate to unload the sample. The images provide unambiguous evidence that the gel was firmly anchored to the two plates, since a gel layer still covers the top (a) and bottom (b) plates.

\begin{figure*}[hb]
\includegraphics[width=12 cm]{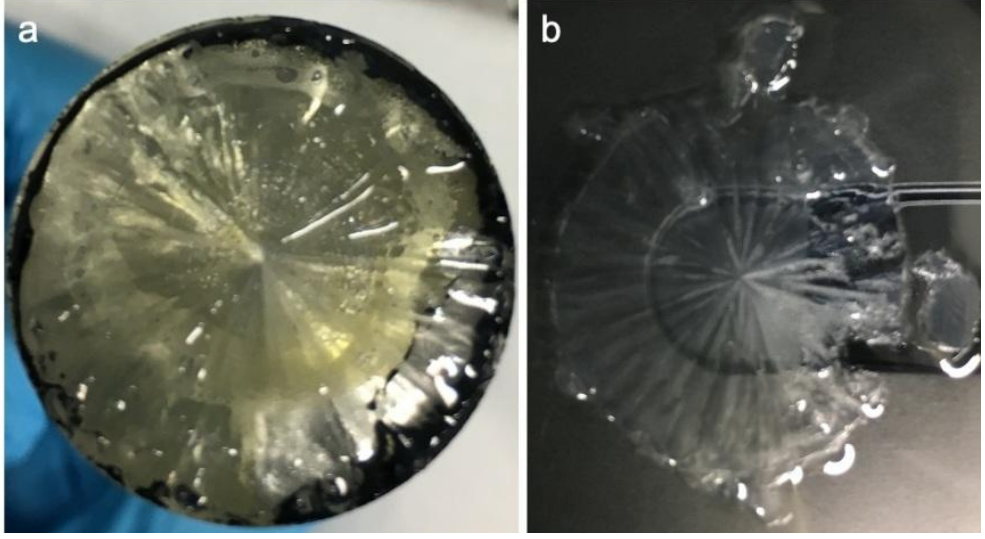}
\caption*{\textbf{Figure SM1:} Images of an agarose gel broken during a creep test. The gel was anchored at the two plates by using polyethyleneimine. When lifting the upper plate to unload the rheometer, the gel brakes in the bulk and does not detach from the plates, as indicated by the gel layer covering both the top (a) and bottom (b) plate.}
\end{figure*}

\textit{Setup.} We show in Fig. SM2 a sketch of the experimental set-up with the location and size of the regions that have been probed by dynamic light scattering in the experiments described in the main manuscript.

\begin{figure}
\includegraphics[width=8 cm]{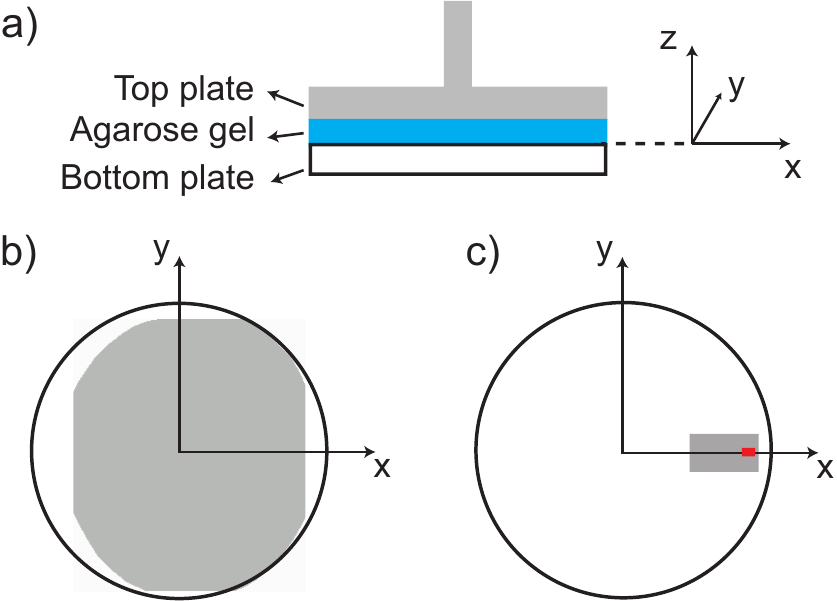}
\caption*{\textbf{Figure SM2:} (a) Side view of the rheometer plates showing the reference system. The $(x,y)$ plane coincides with the interface between the bottom plate and the agarose gel, the $z$ axis is vertical, pointing upward. (b, c) Top view showing the $(x,y)$ plane with the area illuminated by the laser light (grey area) in the two configurations of Fig. 2b and Fig. 3c of the main manuscript, respectively. The red box in (c) indicates the region of interest chosen to investigate the microscopic dynamics of the gel in Fig. 3 of the main manuscript. For the sake of clarity, its size has been enlarged three times with respect to the scale of the rest of the scheme.}
\end{figure}

\section{Computation of the two-time degree of correlation}

The microscopic dynamics are probed in a space- and time-resolved manner by acquiring with a CMOS camera a time series of speckle images, at a rate of 1 Hz. The images are processed to calculate $c_I$, the local, two-time degree of correlation~\cite{duri_resolving_2009,cipelletti_simultaneous_2013}:
\begin{equation} \label{eq:cI}
c_I (t, \tau, \vet{r})  = B \frac{\langle I_p(t)I_p(t+\tau) \rangle_{ROI(\vet{r})}} {\langle I_p(t) \rangle_{ROI(\vet{r})}\langle I_p(t+\tau) \rangle_{ROI(\vet{r})}}-1 \,.
\end{equation}
Here, $\tau$ is a time delay, $B > 1$ a normalization factor chosen such that $c_I$ ($\tau \rightarrow$ 0) = 1, $I_p$($t$) is the time-dependent intensity measured by the $p$-th pixel, and $\langle \cdots \rangle_{\vet{r}}$ is an average over the pixels belonging to a region of interest (ROI) centered around a position $\vet{r}$ in the sample, $\vet{r}=0$ being the coordinate of the rheometer axis of rotation. $c_I$ quantifies the amount of microscopic motion over the time lag $\tau$, averaged over the sample thickness: $c_I \rightarrow 0$ when the scatterers move over a distance $1/q \gtrsim 30$ nm, with $q=33.2~\mu \mathrm{m}^{-1}$ the amplitude of the scattering vector $\vet{q}$, as fixed by the collection optics of the setup~\cite{pommella_coupling_2019}.

\section{Computation of the affine contribution}

During creep, one expects, in addition to possible plastic rearrangements of the scatterers, that the degree of correlation will also decrease because of the motion of the scatterers associated to affine deformation. One can evaluate the amount of correlation that would be observed if the dynamics were due exclusively to the motion of scatterers associated to affine deformation. This contribution can been calculated using the following equation:
\begin{equation} \label{eq:sinc2}
c_{I_{[aff]}} = \text{sinc}^2 \bigg[ \frac{\Delta\gamma(x, y, t, \tau) H} {2}(q_x \sin \beta + q_y \cos \beta) \bigg]
\end{equation}
with $H$ the gap between the rheometer plates, $q_x$ and $q_y$ the $x$ and $y$ component of the scattering vector, $\beta$ the angle between the center of the region of interest (ROI) and the $x$ axis, and $\Delta\gamma(t, \tau)$ the increment of the local strain between times $t$ and $t + \tau$, with $\tau$ = 40 s. In writing Eq.~\ref{eq:sinc2}, we have used the same definition of the reference system as in~\cite{pommella_coupling_2019}: the $(x,y)$ plane coincides with the interface between the bottom plate and the sample and the $z$ axis coincides with the tool rotation axis and is oriented upward. Equation \ref{eq:sinc2} above is the same as Eq. 10 of~\cite{pommella_coupling_2019} in the absence of sources of microscopic dynamics other than the affine deformation (i.e. setting $D=0$ in Eq. 10 of~\cite{pommella_coupling_2019}) and where the strain increment over $\tau$ has been written explicitly, rather than by assuming a constant strain rate as in~\cite{pommella_coupling_2019}.

\section{Additional mechanical and light scattering data}

\subsection{Time evolution of the normal stress during creep tests}

We report in Fig. SM3 the time evolution of the normal stress during the creep test of the four agarose gels reported in Fig. 1 of the main manuscript. We emphasize that upon applying the step of shear stress that starts the creep experiment, the magnitude of the normal stress increases significantly (downward jump of $\sigma_N<0$), as seen by comparing the arrows displaying the normal stress $\sigma_N^0$ just before the step stress to the first value of $\sigma_N(t)$ measured during creep. This jump in normal stress further demonstrates the coupling between shear deformation and normal stress discussed in the main manuscript. The sharp upturn of the normal stress seen for sample S4 (red line) is the signature of the macroscopic failure of the sample.

\begin{figure}
\includegraphics[width=8 cm]{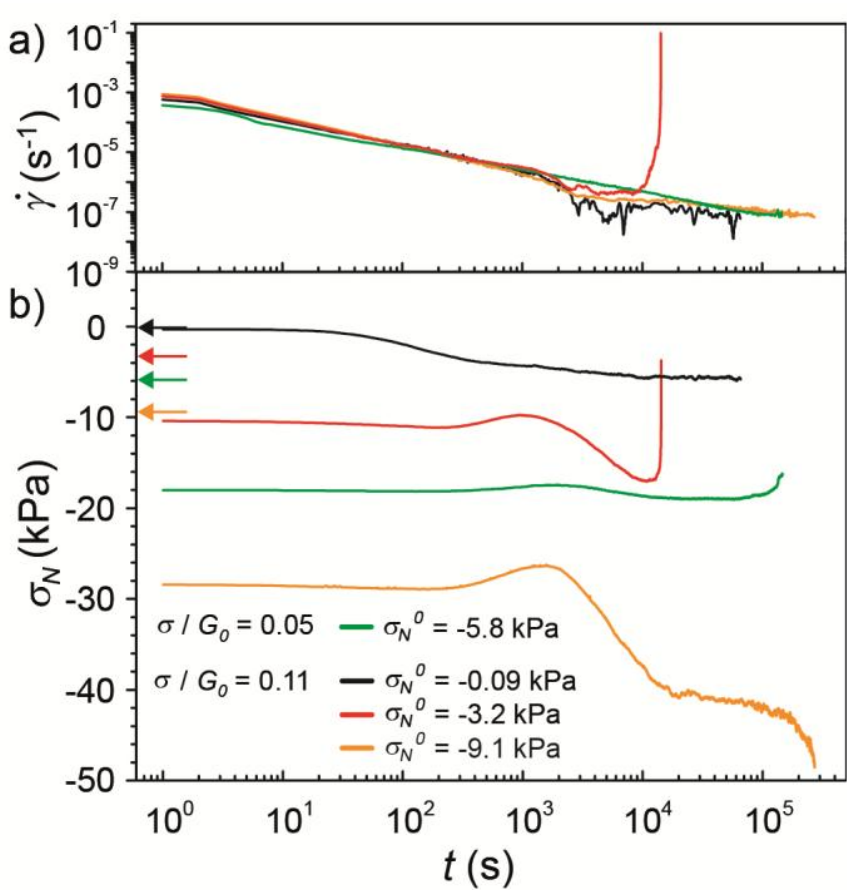}
\caption*{\textbf{Figure SM3:} Time evolution of (a) the shear rate and (b) the normal stress $\sigma_N$ during the creep test of the four agarose gels reported in Fig.1 of the main manuscript. The arrows in (b) show the normal stress $\sigma_N^0$ just before the creep test, when the gels are at rest. For samples S1 to S4, $\sigma_N^0 = -5.8~\mathrm{kPa},-0.09~\mathrm{kPa}-3.2~\mathrm{kPa}$ and $-9.1~\mathrm{kPa}$, respectively. The green curve shows the creep of the gel S1, to which a small stress step $\sigma/G_0 = 0.05$ has been applied. For all the other samples, $\sigma/G_0 = 0.11$.}
\end{figure}

\subsection{Time evolution of the degree of correlation for samples S1-S4}

 We show in Fig. SM4 the time evolution of the degree of correlation $c_I$ for samples S4, the sample that eventually ruptures at $t_R=11825~\mathrm{s}$, during the last $900$ s of the creep. This is the same plot as Fig. 3a of the main manuscript. Here we have added on the graph the affine contribution which represents the amount of correlation that would be observed if the dynamics were due exclusively to the motion of scatterers associated to affine deformation, as computed as described above. The result, $c_{I_{aff}}$, is shown for $\tau = 40~\mathrm{s}$ as an orange line in Fig.~SM4. The $c_{I_{aff}}$ line is close to one and lays above the $c_I$ traces for all the reported time delays, except very close to rupture. This indicates that the lower values seen in the $c_I$ curves for $\tau = 2, 10,$ and $40~\mathrm{s}$ reflect plastic rearrangements leading to enhanced microscopic dynamics, rather than the affine contribution due to the macroscopic deformation.

We show in Fig. SM5 the time evolution of the degree of correlation $c_I$ for samples S1-S4, throughout the whole duration of the creep experiments for sample S4 and during the first $16000$ s for samples S1-S3. Experiments for samples S1-S3 last $24$ hours. Note that, for $t>16000$ s, the degree of correlation $c_I$ for samples S1-S3 are stationary with intermittent small drops (data not shown). For sample S4, after recovering from the initial perturbation, the dynamics significantly accelerate (the degree of correlation $c_I$ drops) between $3500$ and $6000$ s. Subsequently, the dynamics reach a steady state indicative of intense dynamic activity, with a characteristic time scale around $100$ s. Such activity indicates fast internal rearrangements not present for samples S1-S3, the gels that will not rupture, which on the contrary display much longer relaxation time scales (Fig. SM4 a-c). Between $8000$ s and $11000$ s, the dynamics of sample S4 slow down, before a final acceleration sets in on approaching gel rupture. The last $900$ s of the creep of S4 are zoomed in Fig. 3a of the main manuscript and in Fig. SM4, where short time delays ($\tau = 2~\mathrm{s}, 10~\mathrm{s}$ and $40~\mathrm{s}$) are reported to better display these much faster dynamics.

\begin{figure}
\includegraphics[width=12 cm]{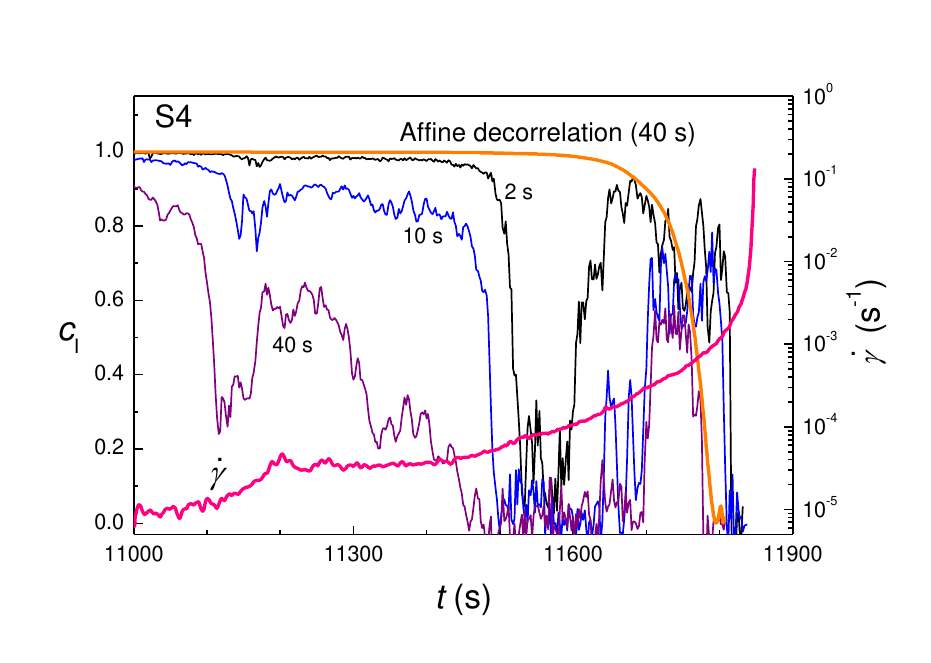}
\caption*{\textbf{Figure SM4:}Time evolution of the degree of correlation at different time delays for the gel showing rupture (S4, red triangles in (b)). Only the last $900$ s before rupture are shown. The macroscopic shear rate (thick pink line) and the affine contribution to $c_I$ for $\tau=40$ s (thick orange line) are plotted. Same figure as in Fig.3 of the main manuscript with in addition the affine contribution to $c_I$.}
\end{figure}

\begin{figure}
\includegraphics[width=12 cm]{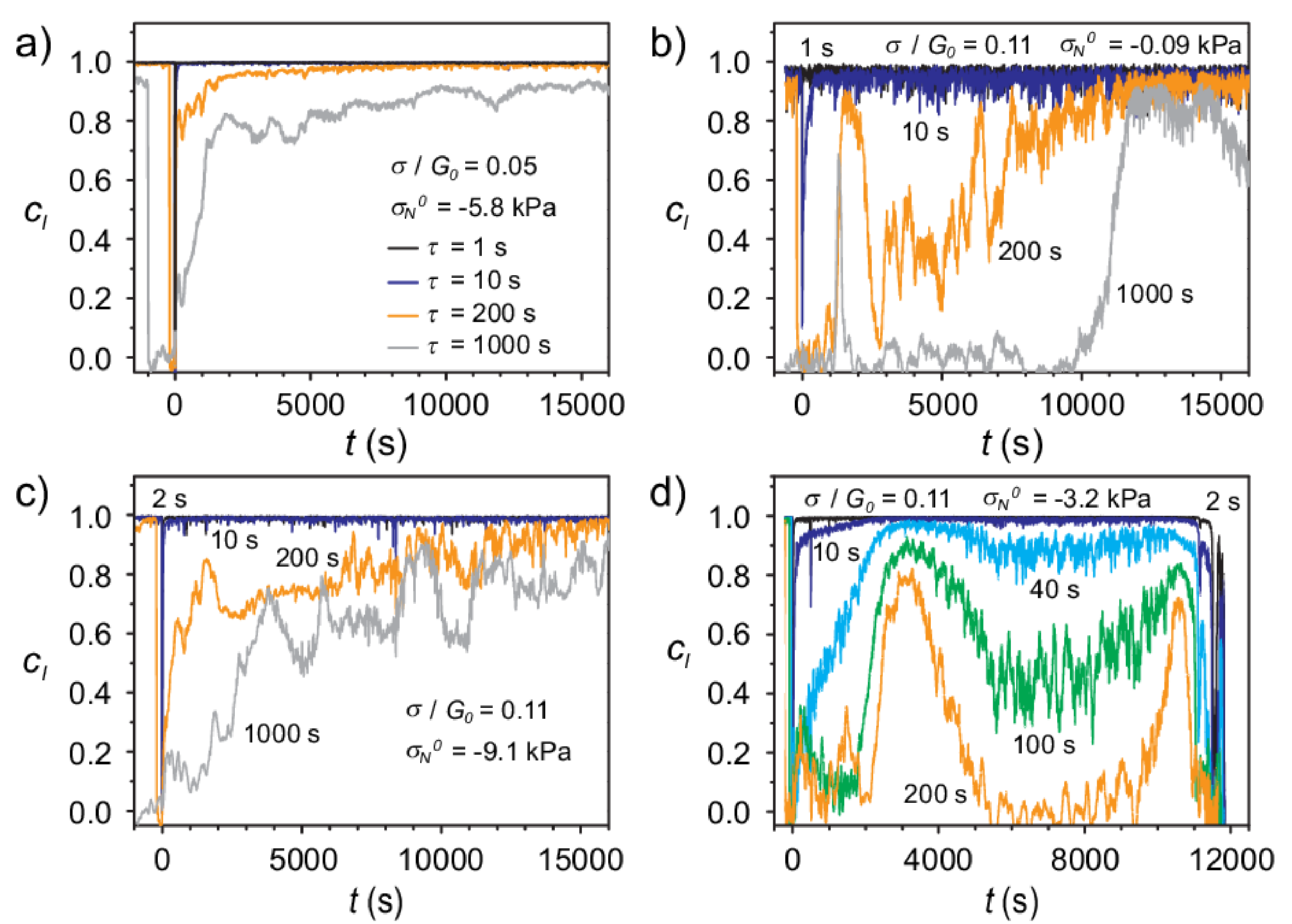}
\caption*{\textbf{Figure SM5:}  Time evolution of the degree of correlation $c_I$ at different time delays $\tau$ during creep tests, for the four samples investigated in the main text: S1 (a), S2 (b), S3 (c), and S4 (d). Time $t=0$ corresponds to the beginning of the creep. The applied stress $\sigma$ and the value $\sigma_N^0$ of the normal stress just before the creep are indicated in the legend. In the creep tests (a-c) no gel failure was seen over $24$ hours. Sample S4 (d) ruptured at $t=11825$ s.}
\end{figure}

\subsection{Intensity correlation functions of gels at rest and during creep experiments}

The fits to the time-averaged degree of correlation with a generalized exponential decay, $g_2 (\tau) - 1 = \exp [-(\tau/\tau_c)^p]$, are shown in Fig. 3b of the main manuscript. We report in Table SMT1 the fit parameters. Note that gels with $\sigma_N^0 \neq 0$  show a compressed exponential decay ($p>1$).  This steeper than exponential decay is due to the presence of internal stresses, as found in a variety of soft solids~\cite{Cipelletti00, Bouchaud01}. By contrast, the gel S1, for which $\sigma_N^0 \approx 0$, shows an exponential decay, indicative of a nearly relaxed initial configuration, free of internal stresses.

\begin{table}
\begin{ruledtabular}
\begin{tabular}{ccccc}
 Sample&$\sigma/G$&$\sigma_N^0$ (kPa)&$\tau_c$ (s)&$p$\\ \hline
 ``at rest''& $0$ & -4.1 & 4300 & 1.9 \\
 S1 & $0.05$ & -5.8 & 2800 & 2.1  \\
 S2 & $0.11$ & -0.09 & 7560 & 0.9  \\
 S3 & $0.11$ & -9.1 & 2550 & 1.5  \\
 S4 & $0.11$ & -3.2 & 140 & 1.4  \\

\end{tabular}
\caption*{Table SMT1. Mechanical parameters and fit parameters for the samples discussed in the text. $\tau_c$ and $p$ are the relaxation time and the stretching exponent of a stretched exponential fit to the intensity correlation function $g_2(\tau)-1$. $\sigma/G$ is the magnitude of the stress step in creep experiments, relative to the sample elastic modulus. $\sigma_N^0$ is the normal stress just before the creep experiment (for S1-S4), or just before measuring $g_2-1$ (for the sample at rest).}
\end{ruledtabular}
\end{table}

\subsection{Cracks at the latest stage of failure}

Just before failure, regions of extremely rapid dynamics develop in the gel. In these regions, the dynamics are so fast that the speckle intensity fluctuates several times during the exposure time (700 ms), resulting in blurred speckles. In these regions the gel is fully fluidized: thus, they correspond to cracks. To visualize them, we calculate, for each pixel, the temporal standard deviation of the speckle intensity $I_p$, analyzing four consecutive images from $17$ s to $13$ s before failure. The result is shown in Fig. SM6, where the standard deviation of $I_p$ is represented in a grey scale, ranging from 0 (black) to $26$ (white) \footnote{As a reference, the spatial standard deviation of a static speckle pattern in the 8-bit images of our experiment is $8.7$ grayscale units.}. The cracks appear as dark zones, originating from the roughly circular region of intense dynamic activity observed a few seconds before and forming lines radially propagating outward. Post-mortem observation of the sample confirms that fractures occur in the bulk, since the two plates are fully covered by a gel layer, similarly to the gel in Fig. SM1.

\begin{figure}
\includegraphics[width=7 cm]{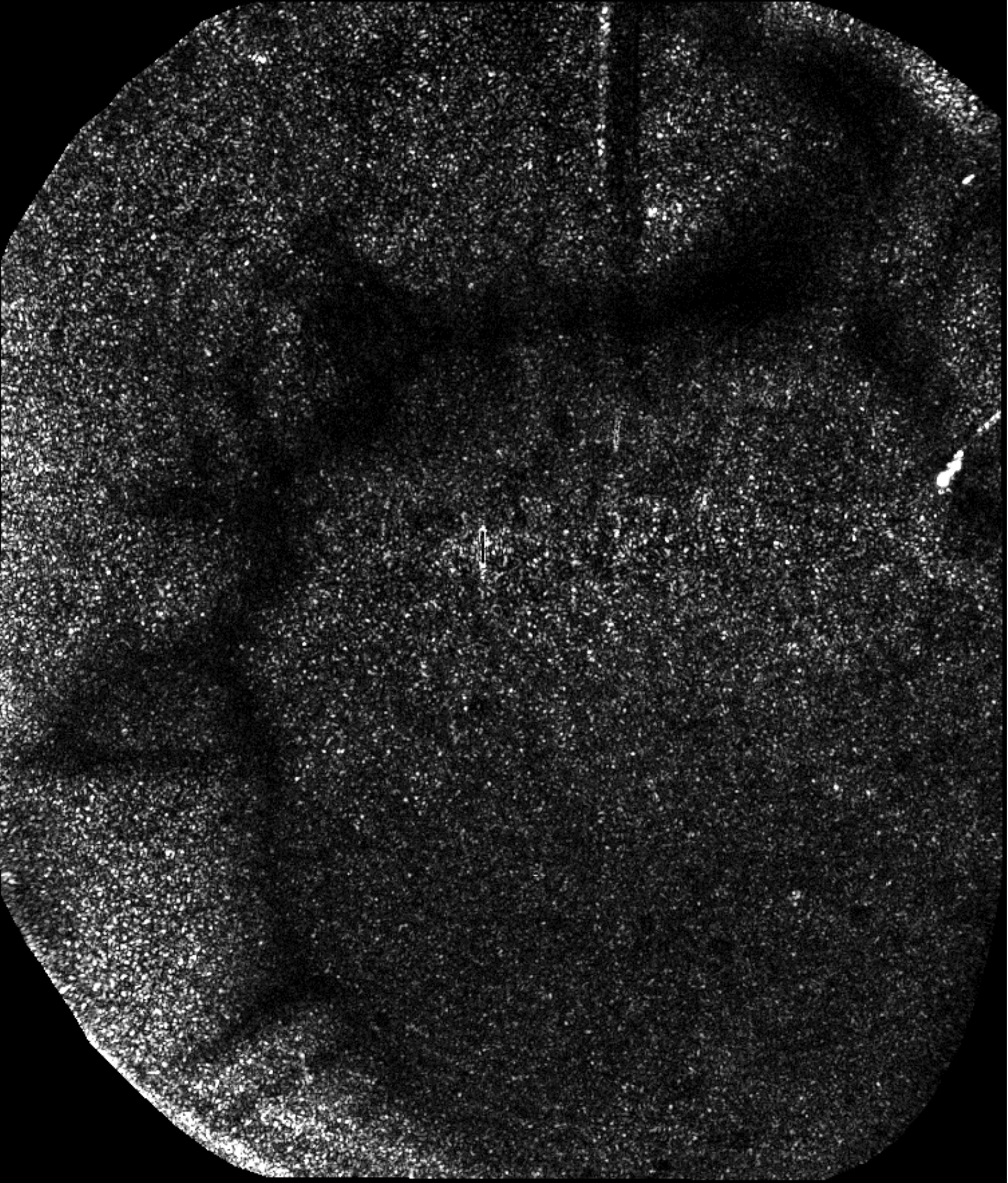}
\caption*{\textbf{Figure SM6:} Standard deviation of the speckle intensity calculated pixel-by-pixel for four consecutive images taken starting 17 s before the gel fails. Same experiment as that for which dynamic activity maps are shown in Fig. 2 of the main text. Cracks originating from a roughly circular region and propagating radially outwards appear as dark zones (see text for details).}
\end{figure}

\subsection{Additional dynamic activity maps}

Figure SM7 b-e display dynamic activity maps (DAMs) measured during several creep experiments as indicated in fig. SM7 a. All DAMs measured with a small or large magnification show a spatially heterogeneous plastic activity before macroscopic failure. Movies of the DAMs for samples S5, S6, S7 and S8 are available as Supplemental materials. The typical size of the early precursors is of the order of a few mm.

\begin{figure}
\includegraphics[width=14 cm]{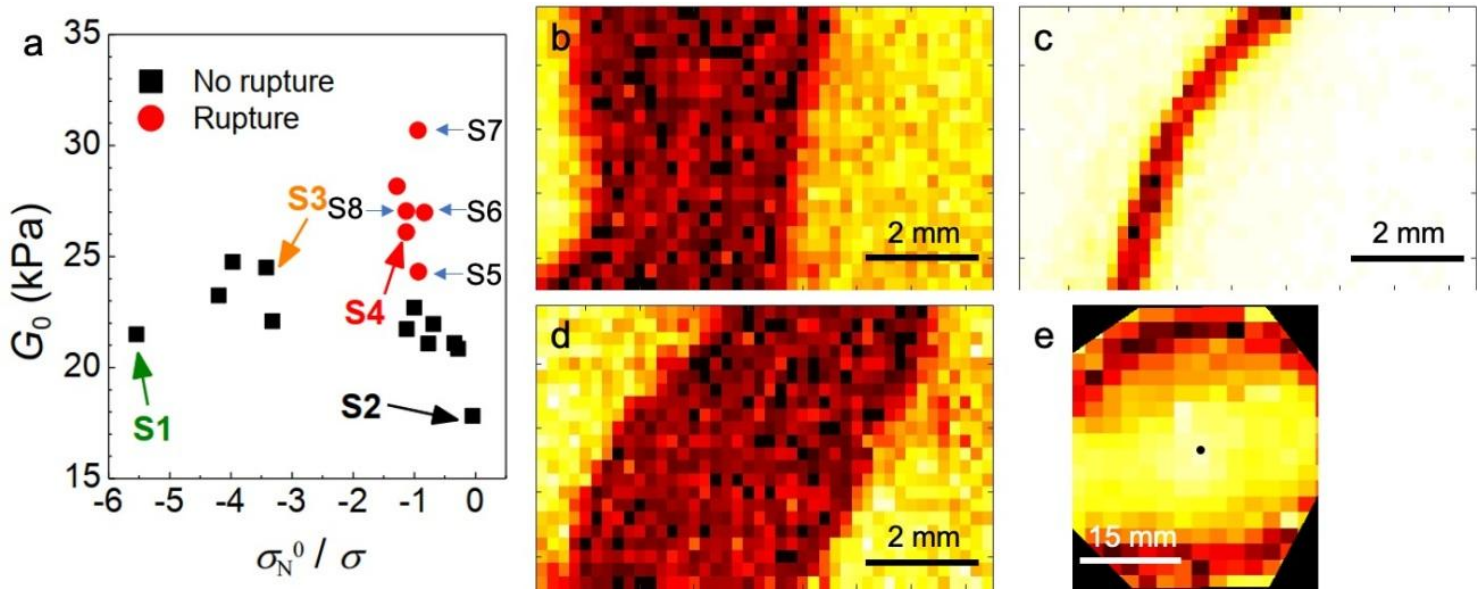}
\caption*{\textbf{Figure SM7:} (a) Elastic modulus as a function of $\sigma_N^0/\sigma$ for gels that eventually did (red circles) or did not (black squares) fail in creep tests. The arrows indicate samples S1-S4 investigated in Figs. 1 and 3 of the main manuscript. The dynamic activity map (DAM) shown in Fig. 2 of the main manuscript was measured for sample S5. The DAMs for the other samples probed by dynamic light scattering (S4, and S6-S8) are comparable to that of samples S5, as shown by panels b) to e) above. (b, c, d): high-magnification DAMs for samples S4 (b), S7 (d) and S8 (c). (e) low-magnification DAM of sample S6.}
\end{figure}

\subsection{Additional rheology data}

Figure SM8 displays the loss modulus and the loss factor measured at $10$ rad/s. For all the gels investigated, the loss factor ($\tan \delta=G”/G’$) is roughly constant: all gels show the same type of response (prevalently elastic, the storage modulus being nearly two orders of magnitude larger than the loss modulus) under deformation independently of the gelation conditions. We find that $\tan \delta$ varies randomly between ca $0.013-0.017$, without showing any trend between yielding and non-yielding gels or with the amount of normal stress (Fig. SM8a). Consequently, we find that the gel loss modulus, $G”$, (plotted as a function of $\sigma_N^0/\sigma$) behaves similarly as the elastic modulus, $G’$ (Fig. 5b in the main manuscript): the gels showing failure are characterized also by higher values of the loss modulus as compared to other gels (Fig. SM8b)

\begin{figure}
\includegraphics[width=14 cm]{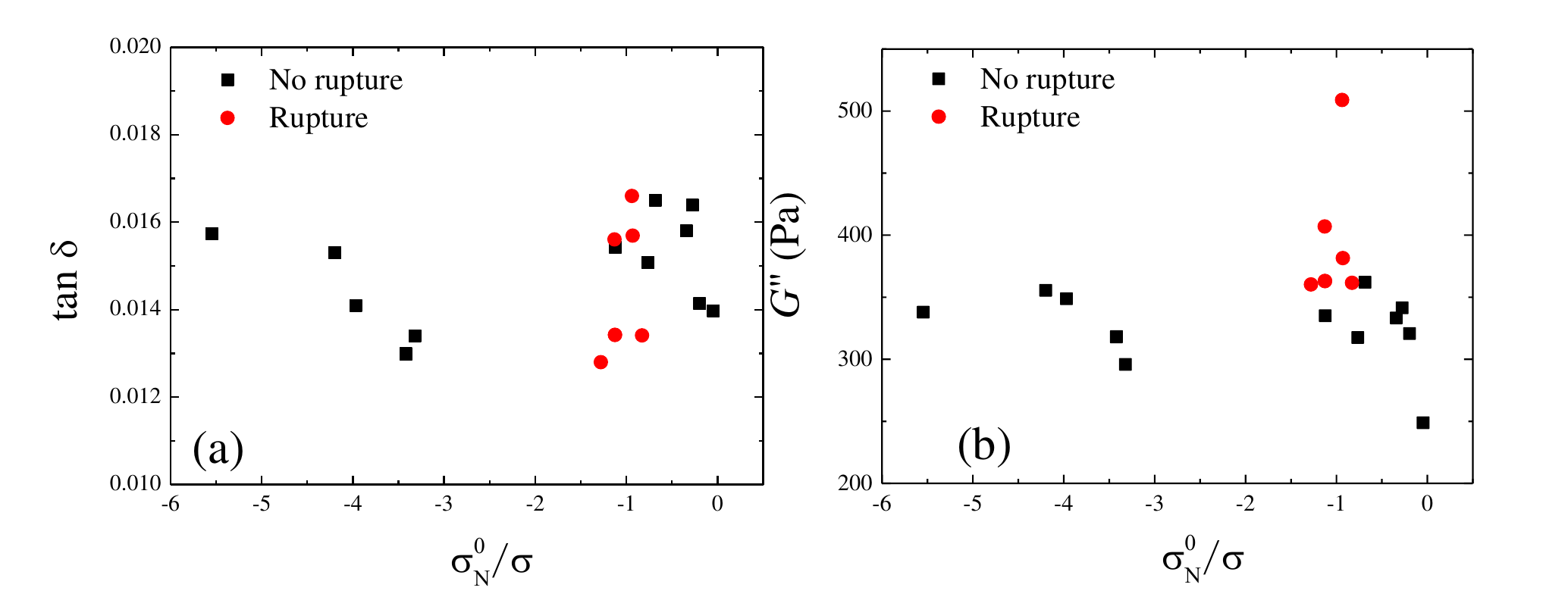}
\caption*{\textbf{Figure SM8:} Loss factor (a) and loss modulus (b) and as a function of $\sigma_N^0/\sigma$ for gels that eventually did (red circles) or did not (black squares) fail in creep tests.}
\end{figure}

\end{widetext}

\end{document}